\documentclass[pre,reprint, doublecolumn,showpacs]{revtex4-1}
\usepackage{amsmath}
\usepackage{graphicx}
\newcommand{\bsym}{\boldsymbol}
\usepackage{ulem}
\usepackage{color}

\begin{document}

\title{Generic Theory of the Dynamic Magnetic Response of  Ferrofluids}

\author{Angbo Fang}
\affiliation{School of Physics and Electronics, North China University of Water Resources and Electric Power, Zhengzhou 450011, China}
\date{\today}

\begin{abstract}
Ferrofluids belong to an important class of highly functional soft matter, benefiting from their magnetically controllable physical properties.
Therefore, it is of central importance to quantitatively predict the dynamic magnetic response of ferrofluids.
Traditional dynamic theories, however, are often restricted to the near-equilibrium regime and/or only apply to nearly ideal ferrofluids that are monodisperse, dilute enough, and weakly interacting.  In this paper I develop a self-consistent and nonperturbative dynamical mean field theory for typical ferrofluids which are often polydisperse, concentrated, and strongly interacting,  possibly driven far from equilibrium.  I obtain a general nonperturbative expression for the dynamic magnetic susceptibility, quantitatively agreeing with the spectra obtained from Brownian Dynamics simulations on both mono- and bidisperse samples. Furthermore,  I derive a generic magnetization relaxation equation (MRE) for both mono- and polydisperse ferrofluids by employing the projection operator technique in nonequlibrium statistical mechanics.  This MRE is in simple closed form and independent of which model is employed to approximate the equilibrium magnetization curve.  Existing models can be recovered as low-order approximations of my generic and nonperturbative MRE.
My theory can play a key role in studying the dynamics of ferrofluids and other polar fluids. It may also have substantial and immediate consequences to various ferrofluid applications.
\end{abstract}
\maketitle
\section{Introduction}
Ferrofluids have been studied extensively~\cite{Rosen:1985, Odenbach:2009} as a prototype of polar fluids and for their many applications in industry and biomedicine.
The number of ferrofluid applications keeps growing, taking advantage of their response to applied magnetic fields which renders their physical (mechanical, optical, thermal, etc.)  properties  controllable and tunable. Thus, it is crucial for us to quantitatively describe the dynamic magnetic response of ferrofluids, which on the macroscopic scale is via a magnetization relaxation equation (MRE).  While the static properties of typical ferrofluids can be reliably predicted by existing models~\cite{Ivanov:1992, Pshenichnikov:1995, HukeLucke:2000, Ivanov:2001, HukeLucke:2004, Ivanov:2007, Ivanov:2017}, however, there still lacks a quantitatively reliable dynamic theory.

The dynamic magnetic response of ferrofluids is mainly complicated by three factors: finite particle concentration, often characterized by the hydrodynamic volume fraction $\phi$; dipole-dipole interactions (DDI) characterized by the coupling constant $\lambda$; and the particle polydispersity.
The earliest dynamic theory of dipolar fluids was developed by Debye~\cite{Debye:1929}, based on the Smoluchowski equation (SE)  for an ensemble of noninteracting dipolar particles (ideal ensemble) under a weak field.  Hence Debye's theory only applies to dipolar fluids on the neighborhood of the unpolarized equilibrium state.  The same restriction applies to phenomenological MREs~\cite{MullerLiu:2001, Felderhof:1999, Shliomis:2001c} obtained from linear irreversible thermodynamics.  By treating the relaxation coefficients as phenomenological state-independent constants, the unjustified use of these MREs in far-from-equilibrium regimes can lead to results even qualitatively wrong.
In 1974,  Martsenyuk, Raikher and Shliomis~\cite{MRSh:1974}(MRSh) proposed an effective field ansatz to derive the MRE for ideal ferrofluids, possibly driven far away from equilibrium by a strong magnetic field or flow deformation.  It can be regarded as a zeroth order approximation (with respect to $\phi \lambda$) to an exact theory and approximately applies to sufficiently dilute and weakly interacting ferrofluids.
However, while there have been some progress in improving the MRSh model via perturbation approaches,  the existing theories are still insufficient for describing dynamic response of real ferrofluids which are often non-dilute, strongly interacting, and polydisperse.

In this paper I develop a generic nonperturbative theory for the dynamic response of  ferrofluids. It quantitatively accounts for the effects of all three aforementioned important issues.  My theory is neither restricted to the near-equilibrium regime nor limited to weakly interacting and dilute ferrofluids.  While on the mesoscopic level it is in essence  a dynamical mean field theory (and can be derived in the framework of the dynamical density functional theory of classical fluids), it proves to be quantitatively reliable as long as particle clustering is insignificant (so that on the macroscopic level it is unnecessary to introduce new structural order parameters coupling to magnetization dynamics).  This implies my theory well captures the time-averaged impact of inter-particle correlations on magnetization dynamics on a slow time scale even though it does not track the much faster evolution of the inter-particle correlation functions themselves.

A generic MRE is analytically derived based on the projection operator method in non-equilibrium statistical mechanics, for
monodisperse as well as polydisperse ferrofluids.   For the first time, we obtain a MRE applicable to polydisperse interacting ferrofluids driven far from equilibrium
(the polydisperse version of the MRSh model is trivial, while other polydisperse theories are restricted to the near-equilibrium regime).
On the other hand, such a MRE, in simple and closed form, can recover all previous MRE's as low-order approximations.  Remarkably, even for a new ferrofluid whose equilibrium magnetization curve is only empirically known but not fitted into any existing analytical models,  this MRE can be readily employed to predict its magnetization dynamics, whether it is near or far from equilibrium.  Importantly, I also obtain a simple, universal, and closed formula for the dynamic magnetic susceptibility (DMS) for both mono- and polydisperse ferrofluids.
Furthermore, I have compared my theoretical predictions with Brownian dynamics simulations.  The quantitative agreements are superior to previous models.  The superiority increases as the ferrofluid samples becomes more concentrated and/or more strongly interacting (see Fig.~1).  Especially, my theory is quantitatively accurate even for  non-dilute interacting ferrofluid samples driven far away from equilibrium (see Fig.~2), for which previous models may even become qualitatively unreliable.

\section{Monodisperse Ferrofluids}
To develop the theory, I first focus on the monodisperse case.
Consider $N$ identical spherical particles (with hydrodynamic diameter $d$, carrying magnetic moment $\mu$) dispersed in a solvent with viscosity $\eta_s$, occupying total volume V, maintained at temperature T, and driven by a magnetic field $\bsym{H}$.  The particle concentration is $\rho = N/V$ and saturation magnetization is $M_s=\rho\mu$.
Typically the relaxation of magnetic moments is dominated by Brownian mechanism. For an ideal ferrofluid ($\phi\sim 0$, $\lambda\sim 0$), particle dynamics is characterized by the Debye or Brownian rotation time~\cite{Debye:1929}
$\tau_D = \pi \eta_s d^3/2 k_B T$ with $k_B$ the Boltzmann constant.  Averaging the single-particle SE for such an ideal ensemble using the effective field ansatz
leads to the celebrated MRSh equation~\cite{MRSh:1974} for magnetization relaxation.
However, for a ferrofluid with finite $\phi=\pi \rho d^3/6$ and $\lambda=\mu_0 \mu^2/4\pi d^3 k_B T$ (with $\mu_0$ the vacuum magnetic permeability thereafter made implicit), particles respond to external fields in a correlated way.  Then we have to solve an intractable N-body problem.  Often we can approximate it by a single-particle SE with an appropriate mean field potential (for a detailed derivation, see Ref.~\cite{Fang:2020x1}).

Denoting $\bsym{e}$ the orientational vector for a representative particle and $W(\bsym{e}, t)$ its orientational distribution function (ODF), the single-particle SE reads
\begin{equation}
2\tau_r \frac{dW}{dt}= \frac{1}{k_B T} \hat{\bsym{J}} W \hat{\bsym{J}}\left[k_B T \ln W - \mu \bsym{e}\cdot \bsym{H} + \Phi_{int}\right],
\end{equation}
with $\hat{\bsym{J}}=\bsym{e}\times \partial/{\partial\bsym{e}}$  and $\tau_r$ the rotational self-diffusion time which is in general larger than $\tau_D$ due to hydrodynamic interactions and DDI~\cite{Cichocki:1999, Nagele:2015}.  Eq.~(1) is similar to the dynamical density functional theory of classical fluids~\cite{Archer:2004}, if we identify
the bracketed term as the total chemical potential $\Phi(\bsym{e}, t) = \Phi_0 + \Phi_{ex} +\Phi_{int}$, with $\Phi_0 = k_B T \ln W$ the ideal gas entropic contribution, $\Phi_{ex}=
-\mu \bsym{e}\cdot \bsym{H}$ the dipole-field potential energy, and $\Phi_{int}$ the effective one-body potential due to DDI.  With Eq.~(1) we can in principle determine the evolution of
magnetization $\bsym{M}(t) \equiv \rho \mu \int \bsym{e} W(\bsym{e}, t) d\bsym{e}$, though it is often difficult to obtain a MRE in closed form.

If $\bsym{H}$ is a static field,  Eq.~(1) admits a stationary solution, $W_0(\bsym{e})$, determined by $\Phi= const$.
Then the magnetic equation of state (MEOS) can be determined, specifying the equilibrium magnetization $\bsym{M}_0$ as a function of $\bsym{H}$.  For example, for an ideal ferrofluid,  $\Phi_0 + \Phi_{ex}= const$ reproduces Langevin's MEOS.  For convenience I define the scaled Langevin function by $\tilde{L}(x)= M_s L(\mu x/k_B T)$, where $L(x)=\coth(x)-1/x$ is the dimensionless Langevin function.  Then Langevin's MEOS is given by $M_0 = \tilde{L}(H)$ and  Langvin's static initial susceptibility is $\chi_L = \rho\mu^2/3 k_B T \equiv 8 \phi \lambda$.
However, taking interparticle correlations  into account  we would have $M_0=\tilde{G}(H)$, with $\tilde{G}$  a function different from $\tilde{L}$.
%
%
%

In Eq.~(1) $\Phi_{int}$ depends on the pair correlation function that usually decays much faster than $W$.   With $\bsym{M}(t)$ the sole relevant slow variable, we may assume $\Phi_{int}$, or the nonequilibrium mean dipolar field $\bsym{H}^{dd}$ defined by $\Phi_{int} = - \mu \bsym{e} \cdot \bsym{H}^{dd}$,  is a functional of $\bsym{M}(t)$ alone.  For example, the Weiss model postulates $\bsym{H}^{dd} = \bsym{M}(t)/3$, from which a MRE can be derived. To the first order of $\chi_L$ it is equivalent to the MRE by Zubarev and Yushkov~\cite{Zubarev:1998, Ilg:2003} for dilute and weakly interacting ferrofluids.  Recently, I have developed a dynamical second-order modified mean field (MMF2)  theory~\cite{Fang:2019b} in which a second-order expression for $\bsym{H}^{dd}$ is constructed.  It applies to ferrofluids with larger $\chi_L$ beyond previous lower-order models.  Now, a general inverse problem arises: Given a magnetization curve $\tilde{G}$ of nonperturbative or even empirical nature,  what is the self-consistent form for $\bsym{H}^{dd}$?

It is a key observation that in all previous perturbative dynamic models~\cite{MRSh:1974, Ilg:2003, Fang:2019b} the mean dipolar field can be obtained from the following universal form:
\begin{equation}
\bsym{H}^{dd}(t) = \bsym{H}^L_e(t) -\bsym{H}_e(t),
\end{equation}
where $\bsym{H}_e(t)=\tilde{G}^{-1}(M(t))\hat{\bsym{m}}$ (with $\hat{\bsym{m}}\equiv\bsym{M}(t)/M(t)$) is the thermodynamic effective field
and  I call $\bsym{H}^L_e(t)=\tilde{L}^{-1}(M(t))\hat{\bsym{m}}$ the (auxiliary) Langevin effective field. With Eq.~(2) the equilibrium ODF is determined as $W_0(\bsym{e}) \propto \exp(\bsym{e}\cdot \bsym{H}^L/k_B T)$, where $\bsym{H}^L = \tilde{L}^{-1}(\tilde{G}(H)) \bsym{H}/H$ is the equilibrium Langevin field.  Thus we have $M_0=\tilde{L}(H^L) \equiv \tilde{G}(H)$, reproducing the prescribed MEOS self-consistently.

I comment that, while Eq.~(2) initially arises as a conjecture based on previous perturbative models~\cite{Fang:2019b}, it can be derived from the dynamical density functional theory of classical fluids with reasonable assumptions.  It should be distinguished from the Weiss mean field model (or other near-equilibrium models) which applies only if the order parameter is sufficiently small (close to the unpolarized equilibrium state), for which we have $\bsym{H}^{dd}(t) \propto \bsym{M}(t)$.  In Eq.~(2) the mean dipolar field is
obtained as a nonlinear function of the instantaneous magnetization. Its explicit form follows self-consistently from the prescribed MEOS, which holds on a slow (quasi-equilibrium) time scale.  Hence, the time-averaged effect of inter-particle correlations  is embodied in my theory to all orders with respect to $M(t)/M_s$.  In contrast, the MRSh model and the Weiss model only captures the correlation effect to the zeroth and first orders, respectively.  This explains why my theory is more generic and accurate, and quantitatively reliable even in the far-from-equilibrium regime.  Furthermore, via Eq.~(2), the mean dipolar field has a physically intuitive interpretation: it is the difference between the magnetic fields needed to prepare the same ferrofluid in a macroscopic quasi-equilibrium state characterized by  $\bsym{M}(t)$, when inter-particle correlations are completely switch off or on, respectively.

To derive the MRE from Eqs.~(1) and (2) I follow Ref.~\cite{Fang:2019b} and employ the projection operator technique~\cite{Zwanzig:1961, Grabert:1982}.
With $\bsym{M}(t)$ the only relevant slow variable, we may discard memory effects and obtain
\begin{equation}
\tau_r \frac{d\bsym{M}}{dt} =\frac{M}{H^L_e} (\bsym{H}-\bsym{H}_e)_{\parallel} + \frac{1}{2}\left(3\chi_L -\frac{M}{H^L_e}\right)(\bsym{H}-\bsym{H}_e)_{\perp},
\end{equation}  
where subscripts ``$\parallel$" or ``$\perp$" denotes the components of $\bsym{H}-\bsym{H}_e$ (the thermodynamic driving force) parallel or perpendicular to $\bsym{M}(t)$, respectively.

 Eq.~(3) is the generic nonperturbative MRE for monodisperse ferrofluids. It is of canonical form~\cite{Blums:1997, Fang:2019a} in nonequilibrium thermodynamics.  Interestingly, the state-dependent transport coefficients are universally determined by the ratio of $M(t)$ to $H^L_e=\tilde{L}^{-1}(M(t))$, its associated instantaneous Langevin field, independent of
 the MEOS.  All previously obtained MRE's, respectively accurate to the zeroth~\cite{MRSh:1974}, first~\cite{Zubarev:1998, Ilg:2003}, and second~\cite{Fang:2019b} order of $\chi_L$,
 can be reproduced by substituting the corresponding MEOS into Eq.~(3) to express $\bsym{H}_e$ as a function of $\bsym{M}(t)$.  Importantly, even if the MEOS  is empirically determined (hence immune to inaccuracies caused by model approximations), we can employ Eq.~(3) to accurately compute the magnetization dynamics.

In some situations $\bsym{H}(t)$ consists of a large static part and a small time-dependent part. If $\mu |\bsym{H}-\bsym{H}_e| \ll k_B T$, then we can
linearize Eq.~(3) and obtain the generalized Debye equation
\begin{equation}
\frac{d\bsym{M}}{dt} = -\frac{(\bsym{M}_{\parallel}-\bsym{M}_H)}{\tau_{\parallel}} - \frac{\bsym{M}_{\perp}}{\tau_{\perp}},
\end{equation}   
where $\bsym{M}_H=\tilde{G}(H(t))\bsym{H}/H$ corresponds to the magnetization for a hypothetical quasi-equilibrium state prepared by the instantaneous magnetic field,
and $\bsym{M}_{\parallel, \perp}$ denotes the components of $\bsym{M}$ parallel or perpendicular to $\bsym{H}$, respectively.
$\tau_{\parallel}$ and $\tau_{\perp}$ are the corresponding field-dependent relaxation times:
\begin{equation}
\frac{\tau_{\parallel}(\xi)}{\tau_r}= \frac{d\xi^L}{d\xi} \frac{d\ln L(\xi^L)}{d\ln \xi^L};\hspace{3 mm}\frac{\tau_{\perp}(\xi)}{\tau_r}= \frac{\xi^L}{\xi} \frac{2 L(\xi^L)}{\xi^L - L(\xi^L)}
\end{equation}  
with $\xi= \mu H/ k_B T$ and $\xi^L = \mu H^L/ k_B T$.

Furthermore, if $\xi\ll 1$,  the magnetization relaxation is restricted around the unpolarized equilibrium and my MRE reduces to a Debye-like equation~\cite{Debye:1929}:
\begin{equation}
\frac{\bsym{dM}}{dt} = -\frac{\bsym{M}-\chi_0 \bsym{H}}{\tau_M} ;  \hspace{4 mm}  \tau_M = \frac{\chi_0}{\chi_L} \tau_r,
\end{equation}   
where $\chi_0$ is the static initial susceptibility.
In Eq.~(6) an exact connection is established between the macroscopic magnetization relaxation time, $\tau_M$,  and the microscopic rotational self-diffusion time, $\tau_r$.  Debye's original equation is for ideal polar fluids with $\chi_0=\chi_L$ and $\tau_M = \tau_r$.  However, for strong ferrofluids we have $\chi_0 \gg \chi_L$ and hence $\tau_M \gg \tau_r$.  Unfortunately, there is a widespread confusion in which $\tau_M$ is improperly identified with $\tau_r$ and the latter is often approximated by $\tau_D$.  Overall this could cause significantly inaccurate estimates of $\tau_M$ or particle sizes,
thereby leading to misinterpretations of relevant experimental measurements.

For noninteracting ferrofluids,  Debye's theory predicts the DMS as $\chi_D(\omega)= \chi_L/(1+ i \omega \tau_r)$.  However, accounting for interaction effects Eq.~(6) leads to
\begin{equation}
\chi(\omega) =  \frac{\chi_0}{1+ i \omega \tau_M}
\end{equation}  
as the generic DMS for monodisperse ferrofluids.  Remarkably, but not obviously,  Eq.~(7) can be recast into the following form:
\begin{equation}
\chi(\omega)= \frac{\chi_D(\omega)}{1 - g_c \chi_D(\omega)/\chi_L};  \hspace{4 mm} g_c\equiv 1 - \frac{\chi_L}{\chi_0}
\end{equation}
where $g_c$ is a measure of interparticle correlations.    Setting $g_c =0$ we have $\chi(\omega)=\chi_D(\omega)$.  If we set $\chi_0 =\chi_L(1+\chi_L/3)$  as in leading-order equilibrium perturbation theories~\cite{Ivanov:1992, Pshenichnikov:1995}, Eq.~(8) exactly reproduces the heuristic modified Weiss model proposed recently~\cite{CampMW:2018}.  If this expression is further subject to a second-order polynomial expansion, then it recovers the DMS from the recent MMF1 theory~\cite{Ivanov:2016a, Note1}.  Thus my theory, of nonperturbative nature and in simple closed form, unifies all previous perturbative results.

\begin{figure}
\centering
\includegraphics [width=1.76 in]{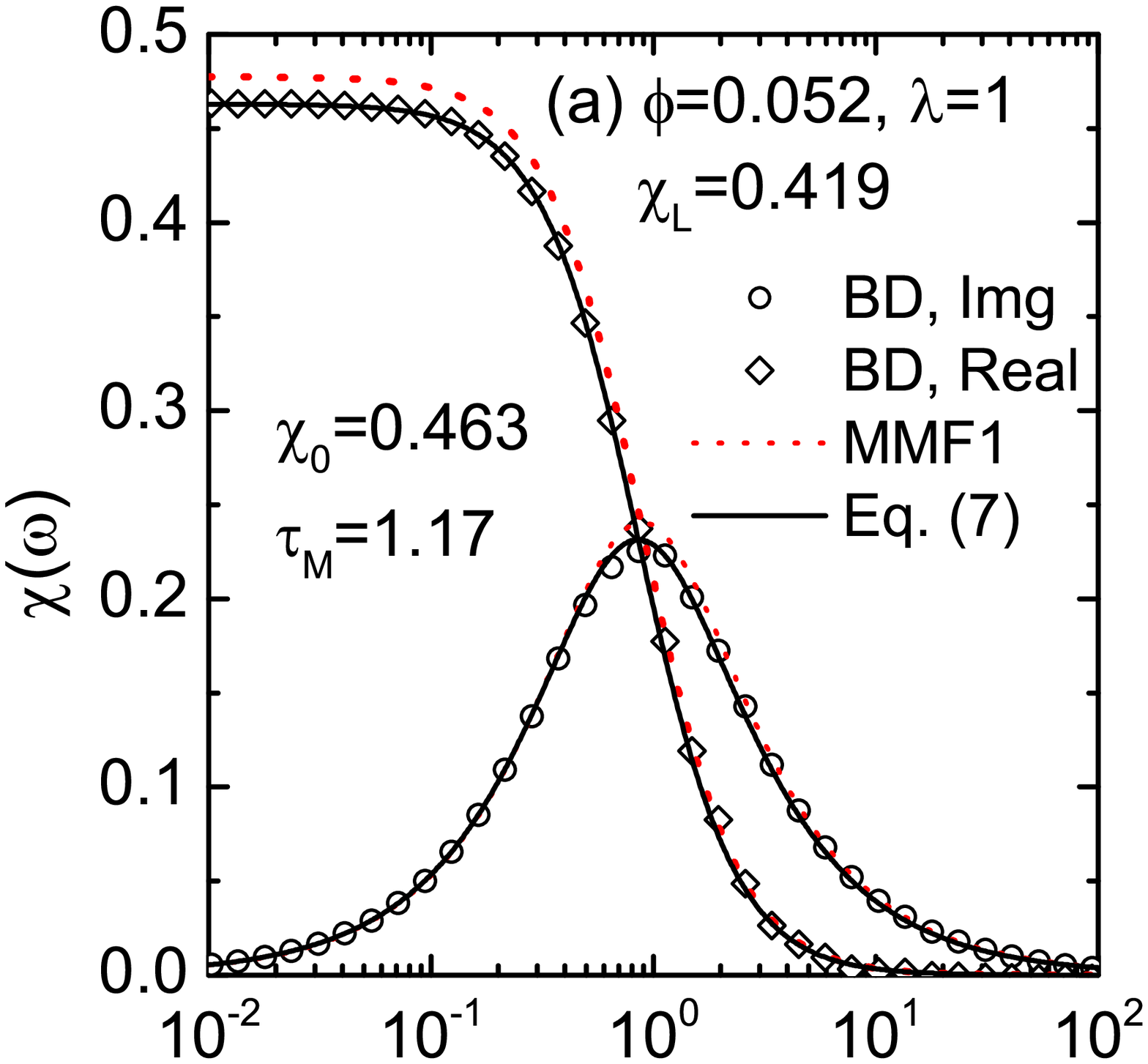}
\includegraphics [width=1.576 in]{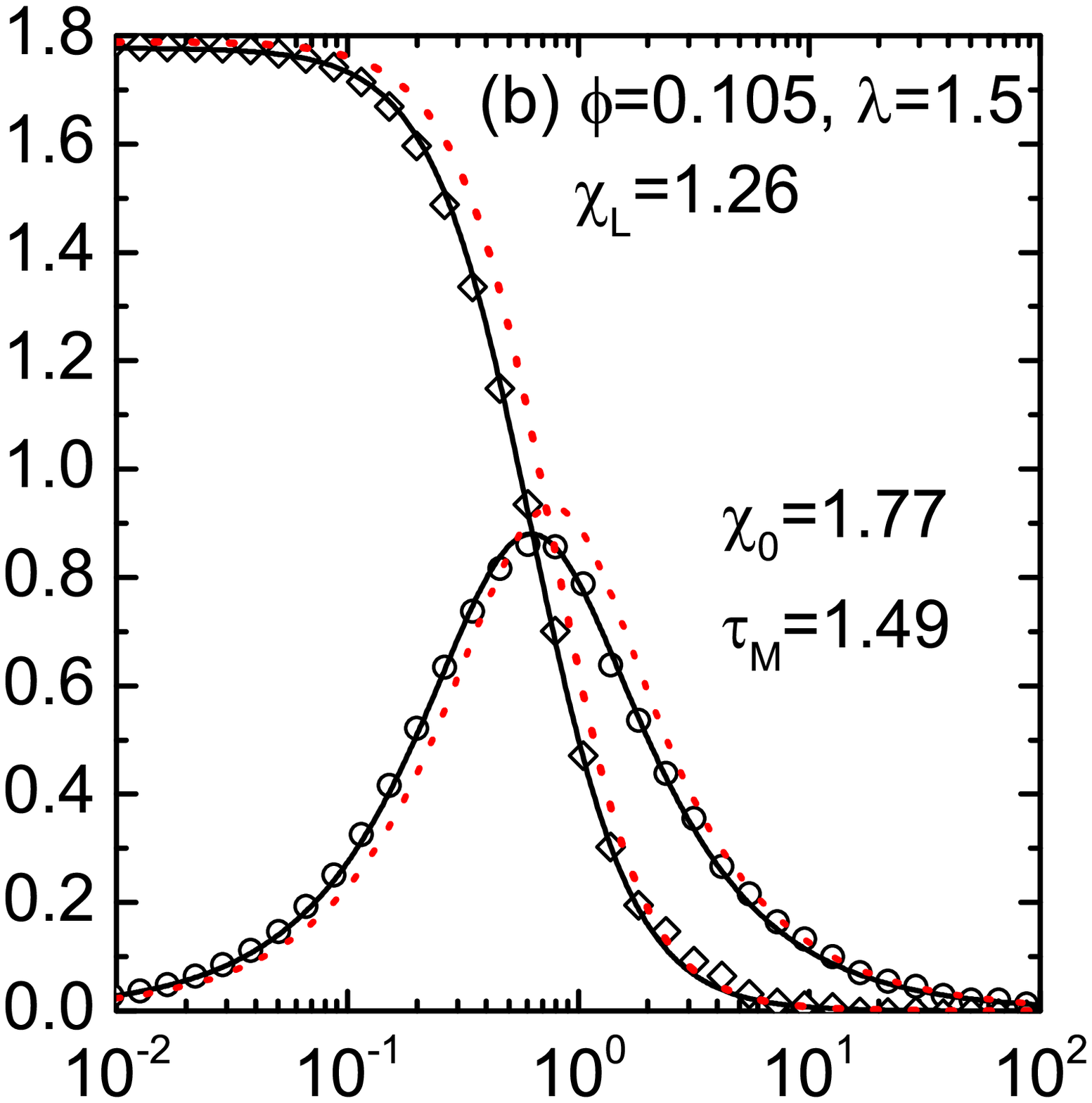}
\includegraphics [width=1.78 in]{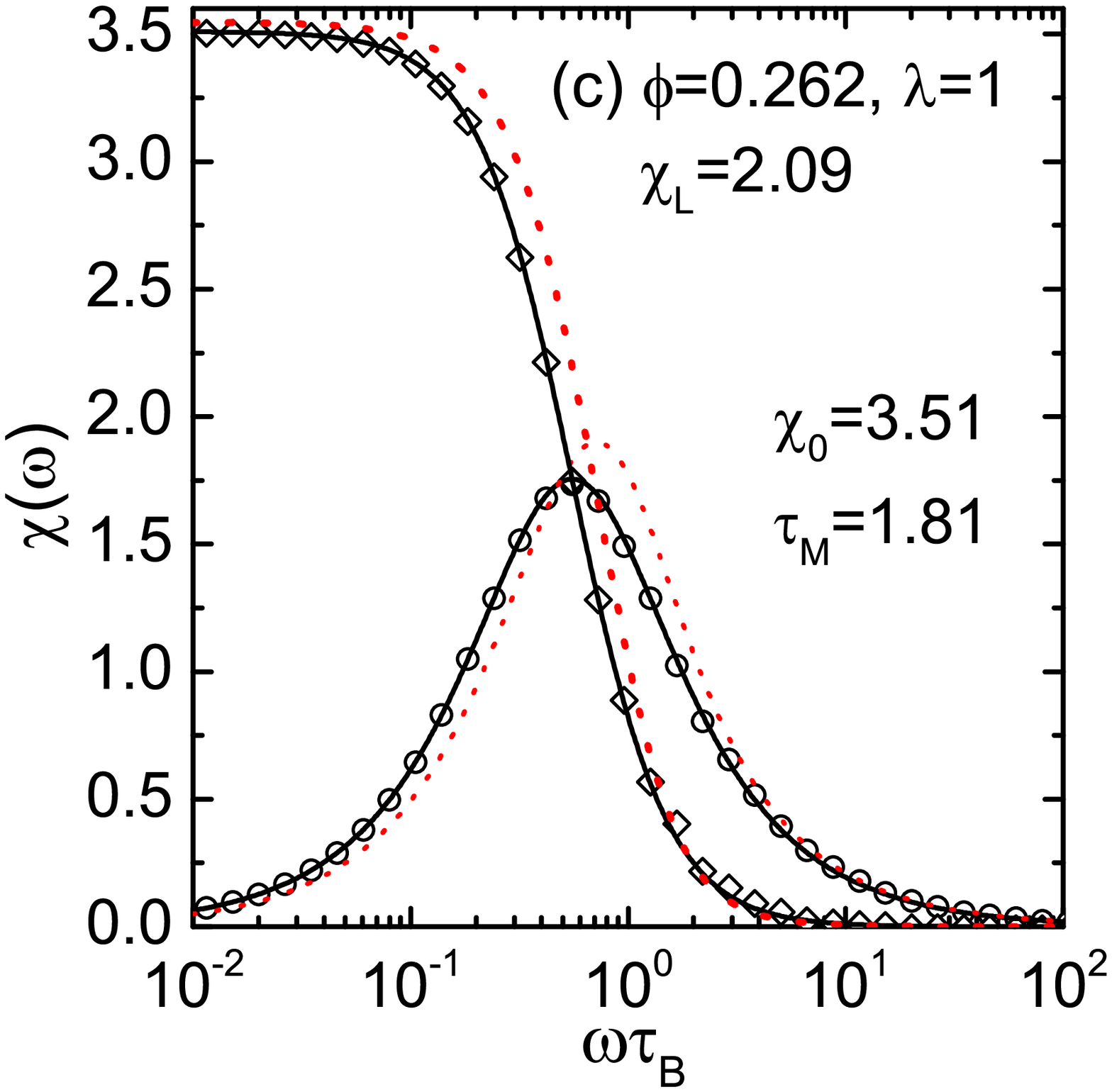}
\includegraphics [width=1.56 in]{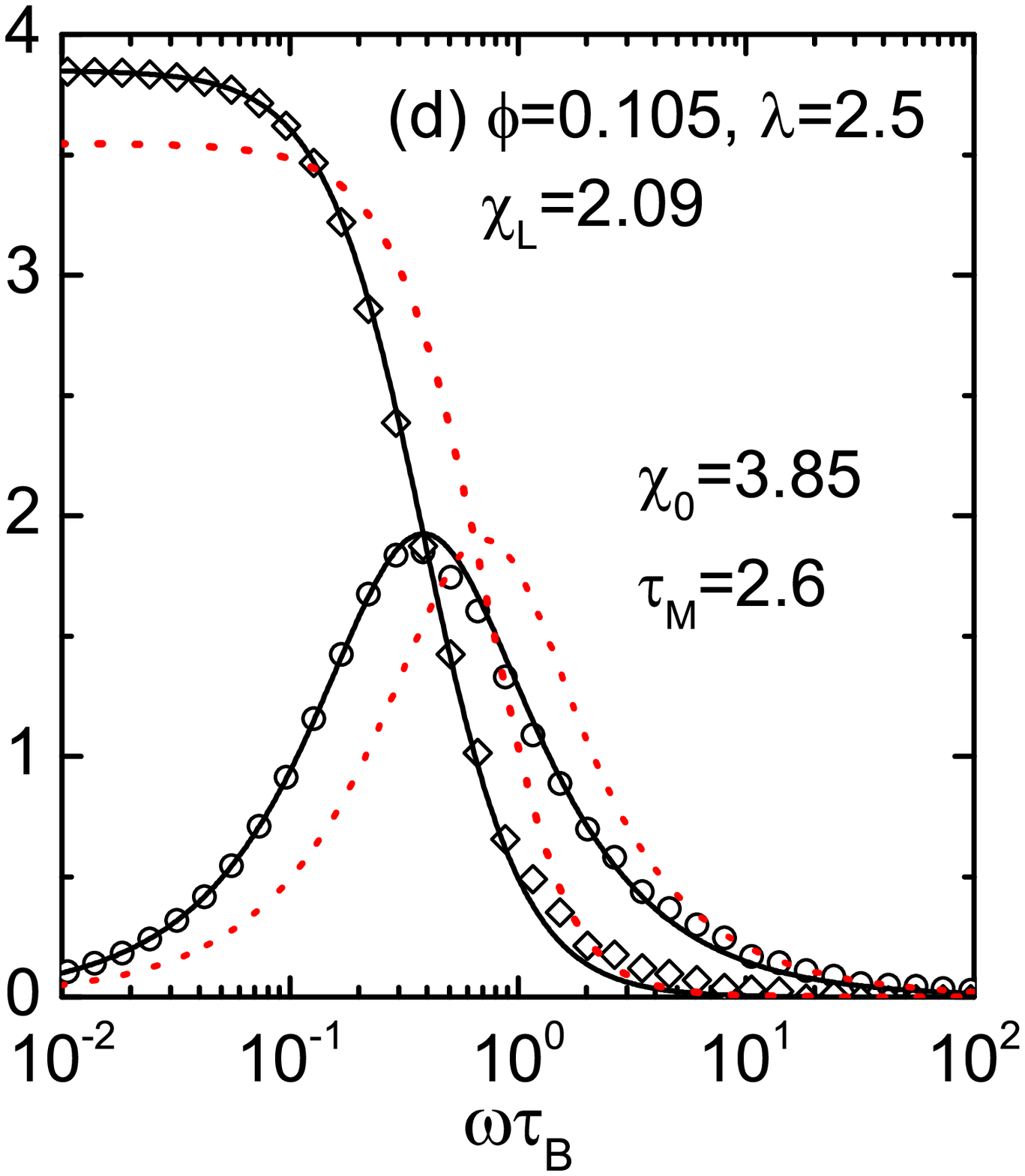}
 \hfill%
\caption {The susceptibility spectra $\chi(\omega)$ for monodisperse ferrofluids with various $\phi$ and $\lambda$.  The symbols are from BD simulations~\cite{Ivanov:2016b} and the dotted lines are from the dynamic MMF1 theory.  The solid lines are from Eq.~(7), with $\chi_0$ and $\tau_M$ (in units of $\tau_B$) specified for each sample.}
\end{figure}
Eq.~(7) can also be directly obtained from the SE (1) with the near-equilibrium approximation to Eq.~(2):
\begin{equation}
\bsym{H}^{dd} = (\chi^{-1}_L -\chi^{-1}_0)\bsym{M}(t).
\end{equation}  
This includes all-order (with respect to $\chi_L$) contributions if $\chi_0$ takes its exact value.  On the other hand, $\bsym{H}^{dd}$ used in all previous DMS calculations ~\cite{Jones:2003, Ivanov:2016a} are at best accurate to the first order of $\chi_L$  and thus fail to describe ferrofluids with either high concentration or strong DDI.  I plot Figure 1 to compare the predicted DMS with the results from Brownian Dynamics (BD) simulations and from the MMF1 theory~\cite{Ivanov:2016b}.  Obviously, the latter becomes less and less accurate as $\phi$ and $\lambda$ increase. In contrast, my predictions agree remarkably well with the simulations for all samples.

\begin{figure}
\centering
\includegraphics [width=1.6 in]{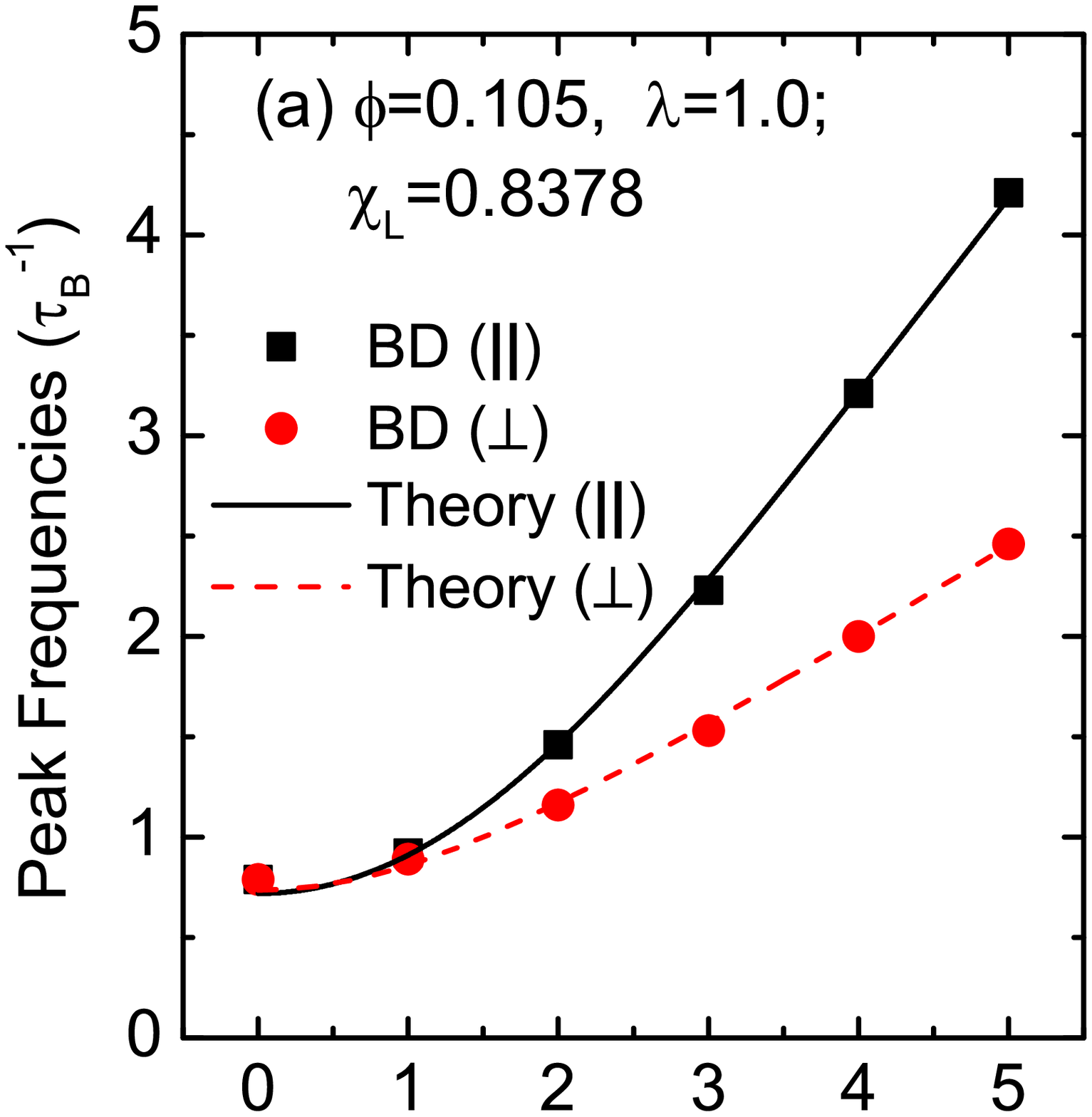}
\includegraphics [width=1.415 in]{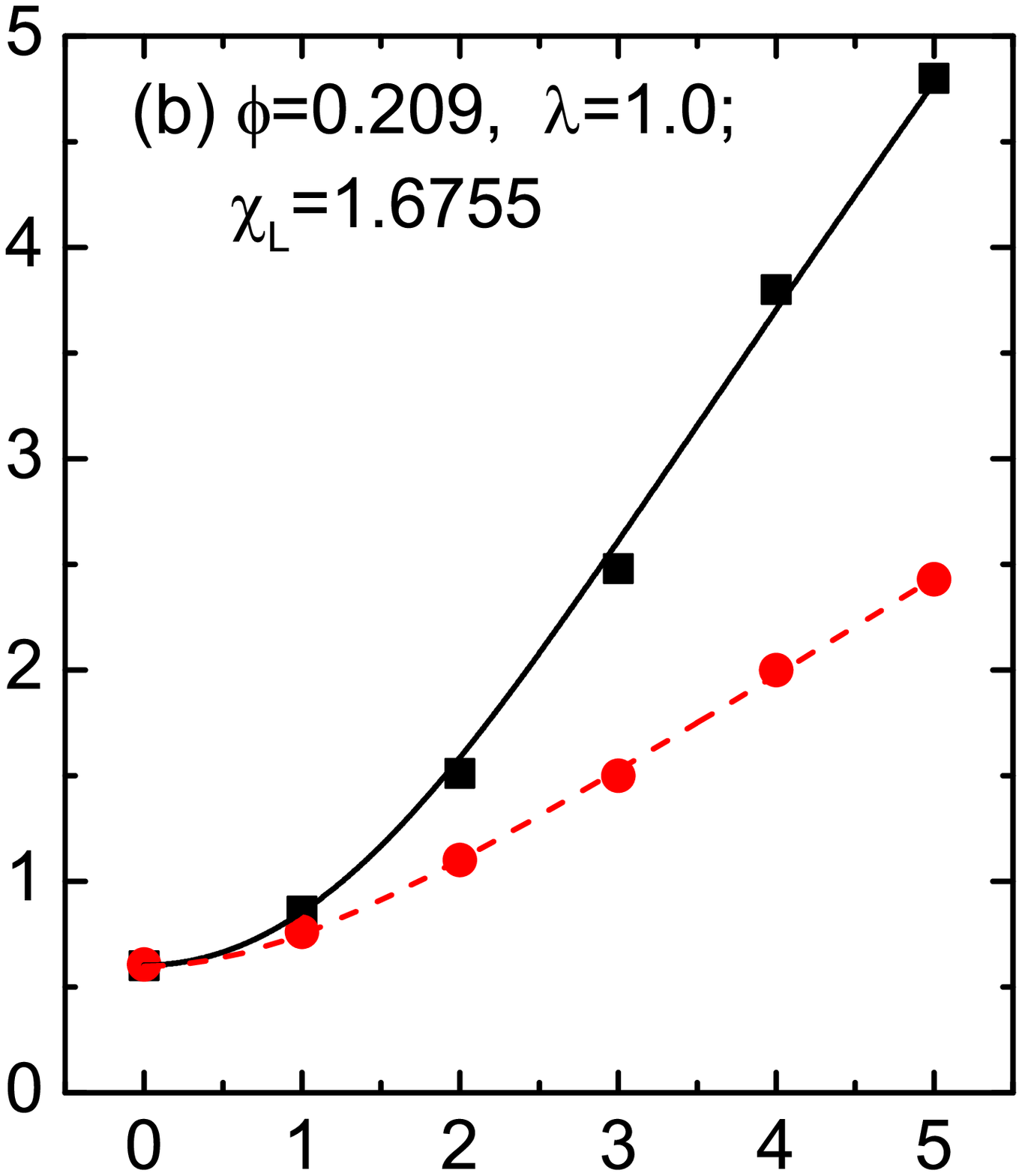}
\includegraphics [width=1.6 in]{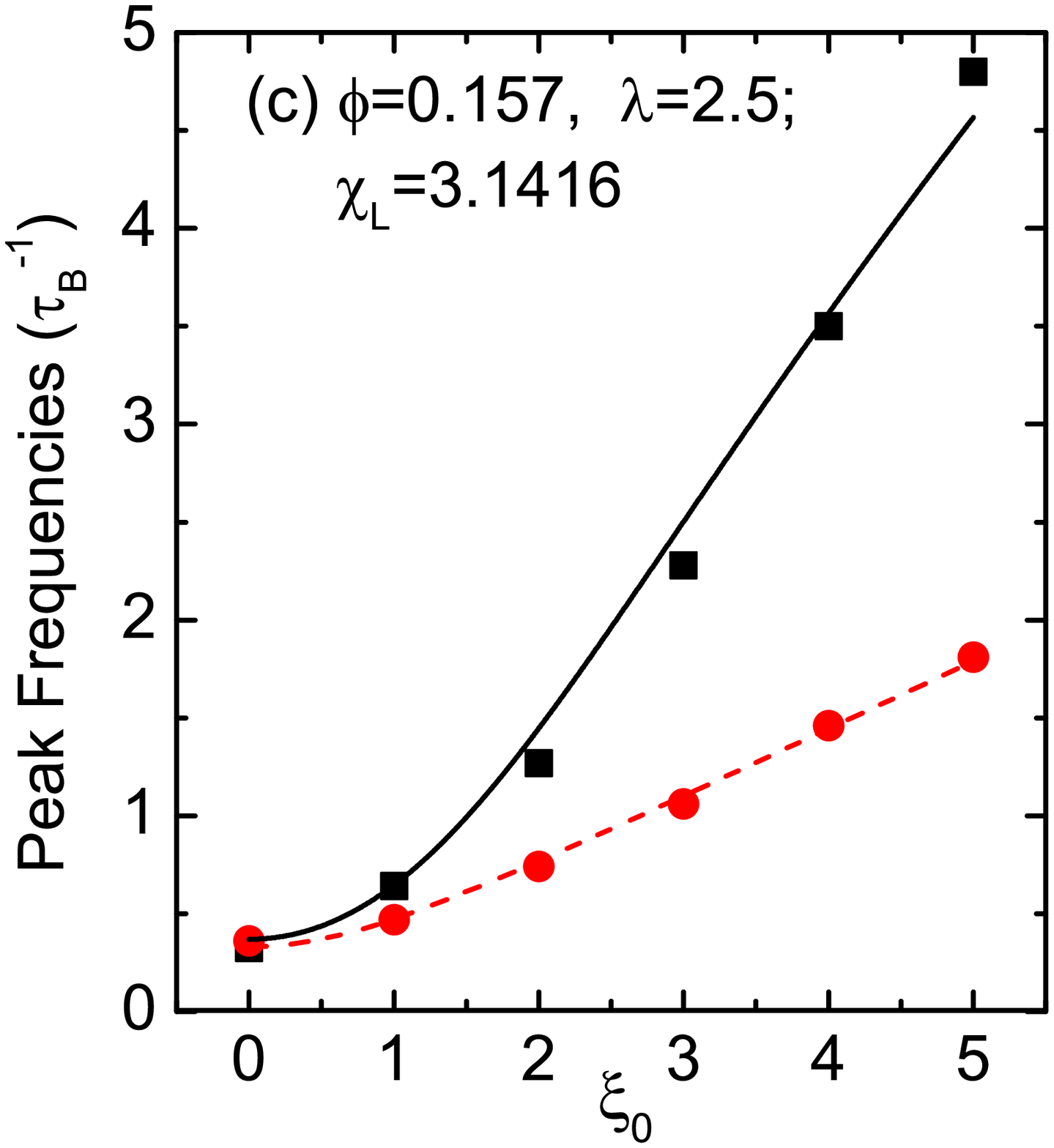}
\includegraphics [width=1.415 in]{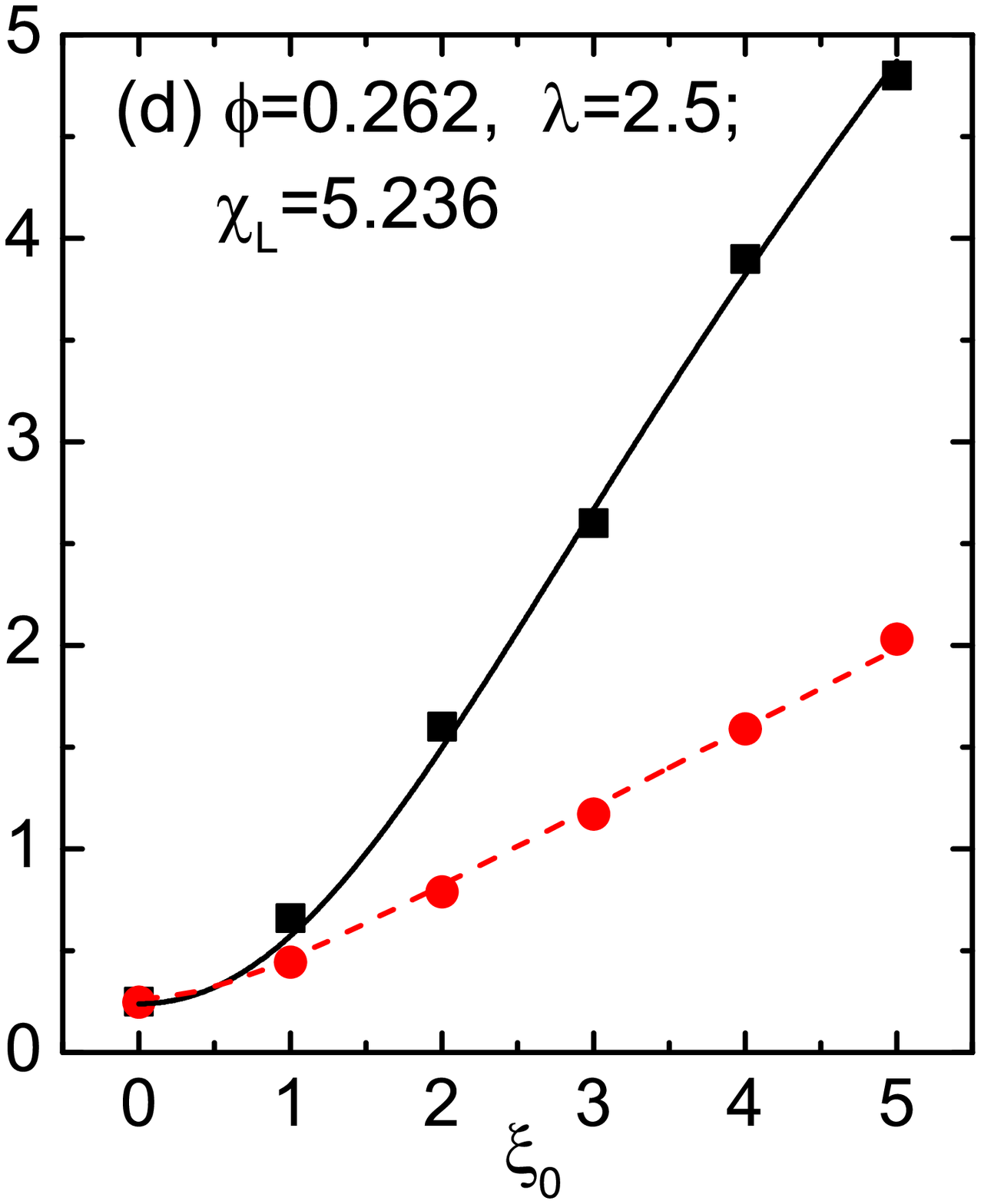}
\includegraphics [width=1.65 in]{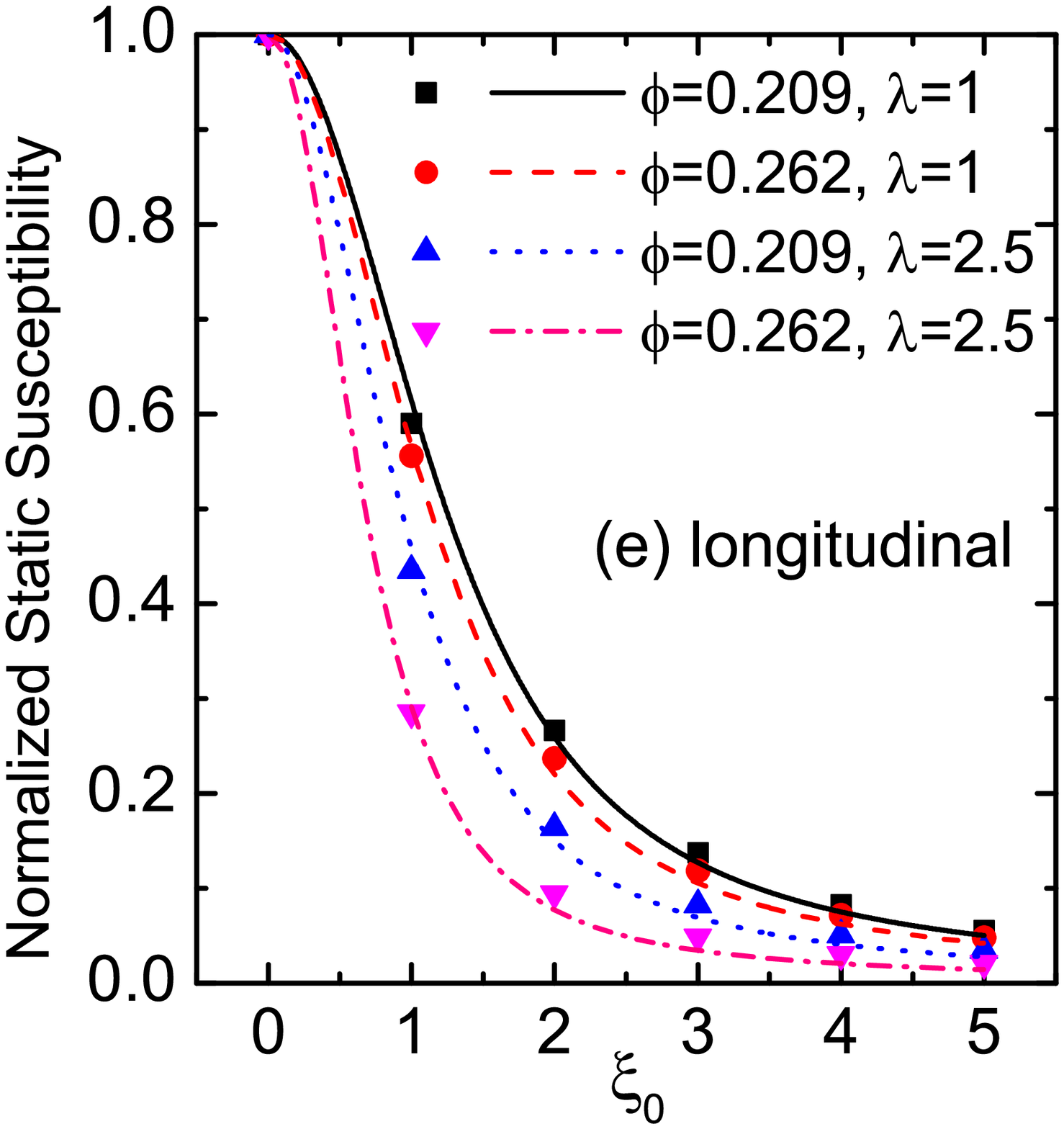}
\includegraphics [width=1.5 in]{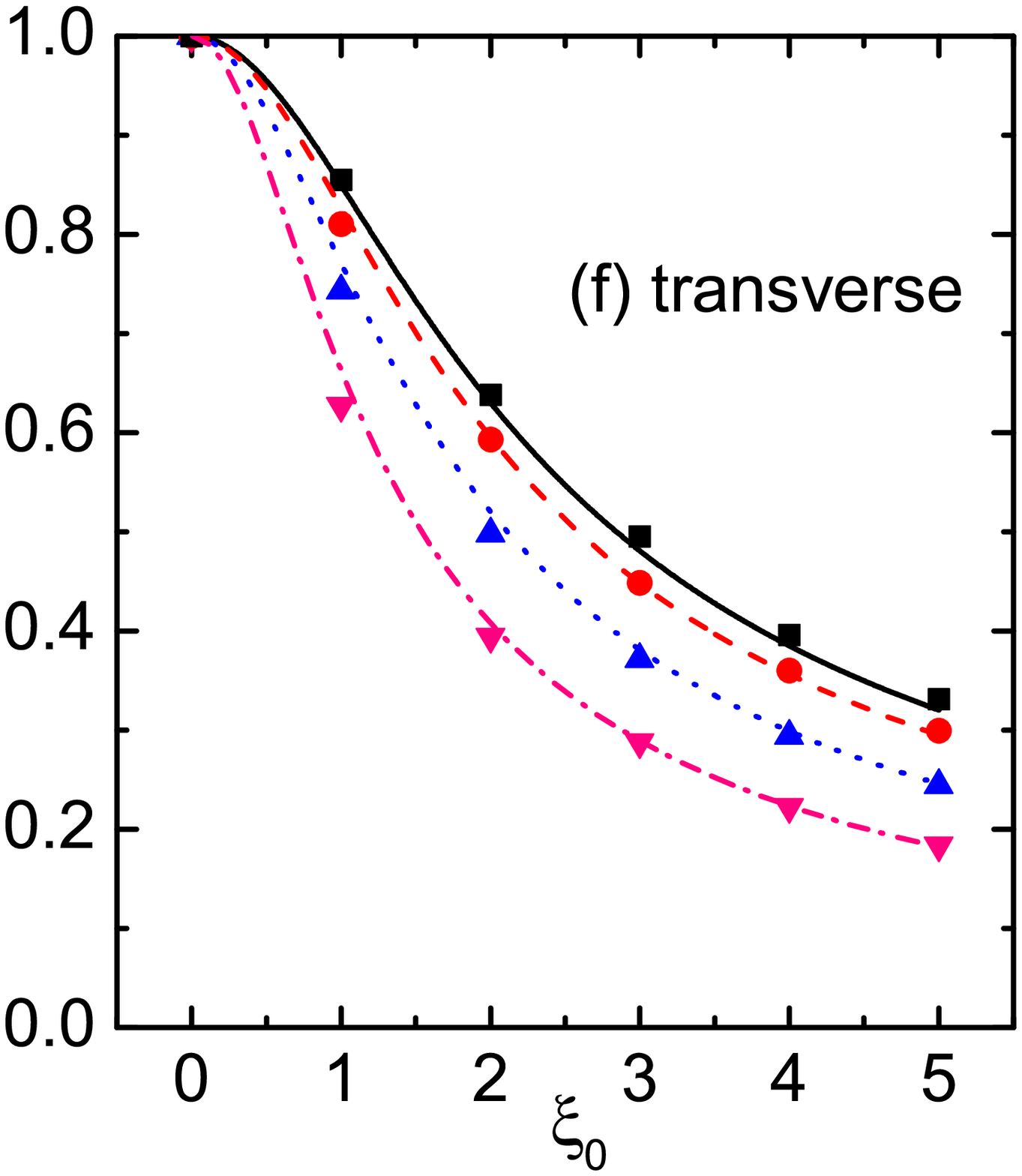}
 \hfill%
\caption {Field dependence of the peak frequencies (a-d) of imaginary-part dynamic susceptibility spectra  and the static susceptibilities (e-f) normalized by their zero-field counterparts, for both the longitudinal ($\parallel$) and transverse ($\perp$) setups. Symbols are from BD simulations in Ref.~\cite{Camp:2018} and lines are from my theoretical predictions. Different samples are distinguished by a pair of values for $\phi$ and $\lambda$.}
\end{figure}

When a ferrofluid sample is put in a static field $\bsym{H}_0$ along with a weak probing AC field $\tilde{\bsym{H}}(t)$, its magnetic response can be described by Eq.~(4).
Unlike previous perturbative models, my theory obtains the finite-field DMS in simple and closed form for both longitudinal ($\tilde{\bsym{H}}\parallel\bsym{H}_0$) and transverse ($\tilde{\bsym{H}}\perp\bsym{H}_0$) setups. The peak frequencies of imaginary-part spectra are respectively given by
$\tau_{\parallel}^{-1}(\xi_0)$ and $\tau_{\perp}^{-1}(\xi_0)$, with $\xi_0=\mu H_0/k_B T$.  Remarkably, my theoretical predictions (supplemented with
MEOS from MMF2 model) quantitatively agree with BD simulation results for all the studied ten samples~\cite{Camp:2018}, with $6\phi/\pi =0.1,...,0.5$ and $\lambda=1 \mbox{  or  } 2.5$.  Figure 2 only shows field dependence of  the static susceptibilities and  peak frequencies for some representative samples. For weakly interacting samples ($\lambda=1$), fitting of peak frequencies according to Eq.~(5) leads to almost the same $\tau_r$ as that determined from the zero-field spectra by Eq.~(7).  For strongly interacting samples ($\lambda=2.5$), $\tau_r$ inferred from the longitudinal and transverse spectra are apparently different.  Only the latter yields a $\tau_r$ close to that from the zero-field spectra (except for the most concentrated sample).  Nevertheless, this properly indicates the formation of transient chain-like structures along $\bsym{H}_0$, which substantially influences relaxation of the longitudinal but not transverse component of magnetization.  Notably, my predictions are even quantitatively good for the sample with $\phi=0.262$ and $\lambda=2.5$ whose MEOS is no longer adequately described by the MMF2 model.  This may indicate the robustness of my theory.

\section{Polydisperse Ferrofluids}
 A real ferrofluid is usually polydisperse and additional complications arise.  For small particles the relaxation of magnetic moments can be dominated by N\'{e}el rather than Brownian mechanism~\cite{Raikher:1994, Vroege:2003, poly1:2016}.  Large-size structural units may form~\cite{Vroege:2003, poly1:2016, poly2:2017, poly3:2018}.  Here,
 I restrict myself to
 typical polydisperse ferrofluids in which the effects due to particle aggregation and N\'{e}el relaxation are of minor importance.

Let us assume $n$ species of particles,  with $p_k$ ($k=1,...,n$) the fraction of particles with magnetic core diameter $d_k$.  The magnetic moment carried by a k-particle is $\mu_k = \pi d^3_k M_s/6$.   I use $\overline{f}$ or $\overline{f_k}$ to denote the population-weighted average of a species-specified quantity $f_k$.  If particle interactions are negligible,  the equilibrium magnetization is given by $\overline{M}_0 = G_0(H)$, with $G_0(x)=\sum_k p_k \tilde{L}_k(x)$ and $\tilde{L}_k(x)= \rho \mu_k L(\mu_k x/k_B T)$.
Langevin's initial susceptibility is $\overline{\chi}_L = \overline{\chi_k^L}$, with $\chi_k^L = \rho \mu_k^2/3 k_B T$.  In  general, as suggested by modified mean field theories, we may write the polydisperse MEOS in the form of $M_0 = G_0(\Lambda(H))$,  with $\Lambda$ a function to be specified. For example, the polydisperse MMF2 model~\cite{Ivanov:2001, Ivanov:2007} postulates $\Lambda(x) = x+ \frac{1}{3} G_0(x) +\frac{1}{144} G_0(x) G'_0(x)$.

Out of equilibrium, I denote $W_k(\bsym{e}, t)$ the k-particle ODF and $\bsym{M}_k(t)$ its first-order moment multiplied by $\rho \mu_k$.  The total magnetization
is $\overline{\bsym{M}}(t)=\sum_k p_k \bsym{M}_k(t)$ and I denote by $\bsym{\overline{m}}$ its director.  I assume $W_k(\bsym{e}, t)$ still satisfies  SE~(1), but with $\tau_r$ replaced by $\tau_k$, the k-particle rotational diffusion time, and $\Phi$ replaced by
\begin{equation}
\Phi_k (\bsym{e}, t) = k_B T \ln W_k - \mu_k \bsym{e}\cdot \left[\bsym{H}+\bsym{H}^{dd}(t)\right],
\end{equation}
the k-particle chemical potential.  In Eq.~(10) the entropy due to species mixing is neglected and $\bsym{H}^{dd}(t)$ is the mean dipolar field. Similar to Eq.~(2),  I postulate
\begin{equation}
\bsym{H}^{dd}(t)= \bsym{H}_e^L(t) -\bsym{H}_e(t),
\end{equation}
where $\bsym{H}_e(t) \equiv \Lambda^{-1}\circ G_0^{-1}\left(\overline{M}(t)\right) \overline{\bsym{m}}$ and $\bsym{H}^L_e(t) \equiv G_0^{-1}\left(\overline{M}(t)\right) \overline{\bsym{m}}$ are respectively the thermodynamic and Langevin effective field associated with $\overline{\bsym{M}}(t)$.  Hence, a particle, to whichever species it belongs, is subjected to the same mean dipolar field (with inter-particle correlations time-averaged).  In general, we expect Eq.~(11) is only a good approximation on a sufficiently slow time scale when $\overline{\bsym{M}}(t)$ is the sole relevant slow variable.  However, when the polydisperse ferrofluid remains close to the unpolarized equilibrium state (for which Eq.~(11) can be linearized with respect to $\overline{M}(t)$), it may also perform well on a time scale faster than the time required to equilibrate all $\bsym{M}_k$ in the far-from-equilibrium regime.  The quantitative reliability of Eq.~(11) should be checked {\sl a posterior}.

Now, treating $\{\bsym{M}_k; k=1, ..., n\}$ as relevant slow variables,   we can manipulate the k-particle SE  using the projection operator technique and obtain
\[\tau_k\frac{d\bsym{M}_k}{dt} = \frac{M_k}{H^L_e} (\bsym{H}-\bsym{H}_e)_{\parallel} +\frac{1}{2}\left [3 \chi^L_k -\frac{M_k}{H^L_e}\right](\bsym{H}-\bsym{H}_e)_{\perp}\]
\begin{equation}
 + \int_0^t  ds \hspace{1 mm} {\mathcal{A}}_k(\bsym{M}_k(s), \overline{\bsym{M}}(s), t, s) [\bsym{H}(s)-\bsym{H}_e(s)],
\end{equation}
where on the right hand side the first and second line represents the instantaneous response and memory effects, respectively. The memory kernel ${\mathcal{A}}_k$ is a second-order tensor and depends on  all $\bsym{M}_k$ in the past.  Unlike the monodisperse case with only one relevant slow variable, here we can not discard the memory effects.  This is because the relaxation times for different species can be drastically different and their interactions can not be
adequately described by instantaneous coarse-grained forces.

Still, on a time scale slow enough to wash out memory effects and validate the quasi-equilibrium approximation, a polydisperse ferrofluid should also obey the principle of nonequilibrium thermodynamics. The rate of change of $\overline{\bsym{M}}(t)$ should be proportional to the  instantaneous nonequilibrium thermodynamic force $\bsym{H}(t)-\bsym{H}_e(t)$ with transport coefficients depending solely on $\overline{\bsym{M}}(t)$, irrespective of whether the ferrofluid sample is monodisperse or polydisperse.  Therefore, it is expected that the memory effects are  to regulate the relaxation rates of different species so that all $\bsym{M}_k (t)$ become synchronized, rendering $\overline{\bsym{M}}(t)$ the sole and adequate slow variable characterizing the instantaneous thermodynamic state.  Hence we may absorb the memory-effect term into $\tau_ k \frac{d\bsym{M}_k}{dt}$ and write
\begin{equation}
\overline{\tau}_B \frac{d\bsym{M}_k}{dt} = \frac{M_k}{H^L_e} (\bsym{H}-\bsym{H}_e)_{\parallel} +\frac{1}{2}\left [3 \chi^L_k -\frac{M_k}{H^L_e}\right](\bsym{H}-\bsym{H}_e)_{\perp},
\end{equation}
in which $\overline{\tau}_B$ could be a function of $\overline{\bsym{M}}$.

Taking the population-weighted average of Eq.~(13) leads to the polydisperse MRE
\begin{equation}
\overline{\tau}_B \frac{d\overline{\bsym{M}}}{dt}=\frac{\overline{M}}{H^L_e} (\bsym{H}-\bsym{H}_e)_{\parallel} +\frac{1}{2}\left [3 \overline{\chi}_L -\frac{\overline{M}}{H^L_e}\right](\bsym{H}-\bsym{H}_e)_{\perp}.
\end{equation}
For Eq.~(14) to reduce to Eq.~(3) in the monodisperse limit, $\overline{\tau}_B$ should be state independent.  Therefore,  $\overline{\tau}_B$ can be identified as the averaged rotational diffusion time.

Near the unpolarized equilibrium Eq.~(14) reduces to a Debye-like equation and a DMS can be deduced:
\begin{equation}
\overline{\chi}_s (\omega) = \overline{\chi}_0 / (1+ i \omega \overline{\tau}_B \overline{\chi}_0/\overline{\chi}_L),
\end{equation}
with $\overline{\chi}_0$ the static susceptibility. Nevertheless, $\overline{\chi}_s(\omega)$ is only a low-frequency approximation to the true DMS.
This is distinguished from monodisperse ferrofluids, for which the same DMS [Eq.~(7)] is derived from either the macroscopic MRE [Eq.~(3)] or microscopic (strictly speaking, mesoscopic) model [Eq.(9)].  This is because for monodisperse ferrofluids there is a single microscopic time scale and the magnetization is the unique slow variable. However, for polydisperse ferrofluids, there are multiple microscopic time scales and multiple mutually coupled slow variables. Except at a sufficiently slow time scale, memory effects can not be discarded, even in the linear response regime. Hence we have to go back to the SE to obtain the polydisperse DMS.

I denote the weak AC magnetic field as $\bsym{H}(t) =\bsym{H}_0 \exp(i \omega t)$, with $\xi^0_k=\mu_k H_0/k_B T \ll 1$ for $k=1,..., n$.  Under the linear response approximation we have $\bsym{M}_k (t) = \chi_k(\omega) \bsym{H}(t)$ and $\overline{\bsym{M}} (t) = \overline{\chi}(\omega) \bsym{H}(t)$  with $\overline{\chi}(\omega) = \sum_k p_k \chi_k (\omega)$.
To the first order of $\xi^0_k$, the mean dipolar field becomes
\begin{equation}
H^{dd} = \left(\overline{\chi}^{-1}_L -\overline{\chi}^{-1}_0\right)\chi(\omega) \bsym{H}_0 \exp(i \omega t)
\end{equation}
and the k-particle ODF is
\begin{equation}
W_k(\bsym{e}, t) = 1 + \frac{\chi_k(\omega)}{\chi^k_L} \frac{\mu_k \bsym{e}\cdot\bsym{H}_0}{k_B T}\exp(i \omega t).
\end{equation}
Plugging Eqs.~(16) and (17) into the SE's, we obtain
\begin{equation}
(1+i \omega \tau_k ) \chi_k(\omega) = \chi^L_k \left[1+ (\overline{\chi}_L^{-1}-\overline{\chi}_0^{-1}) \overline{\chi}(\omega)\right]
\end{equation}

The population-weighted average of Eq.~(18) reads
\begin{equation}
\overline{\chi}(\omega)= \frac{\overline{\chi}_D(\omega)}{1 - g_c \overline{\chi}_D(\omega)/\overline{\chi}_L}
\end{equation}
with $g_c = 1- \overline{\chi}_L/\overline{\chi}_0$ the correlation factor and
\begin{equation}
\overline{\chi}_D = \sum_k p_k\chi^L_k /(1+ i \omega \tau_k)
\end{equation}
the Debye's DMS for noninteracting ploydisperse ferrofluids.   Remarkably, Eq.~(19) is completely parallel to Eq.~(8) for the monodisperse DMS.  In both cases, the correction to Debye's DMS only depends on a single correlation factor $g_c$.  This correlation factor only depends on the static initial susceptibility, which can be determined either empirically or by approximate equilibrium models.  If my theory is supplemented with $\overline{\chi}_0 = \overline{\chi}_L (1+ \overline{\chi}_L/3)$ as
predicted by first-order equilibrium perturbation theories and neglect the difference between the $\tau_k$ and $\tau_{Bk}$ (which denotes the Brownian relaxation time
for the k-particle in the infinitely dilute limit),  then Eq.~(19) reduces to the heuristic modified Weiss model~\cite{CampMW:2018}.
The latter was shown in good agreement with BD simulations for several bidisperse samples with $\overline{\chi}_L \le 3.2$.

\begin{figure}
\centering
\includegraphics [width=4.4 in]{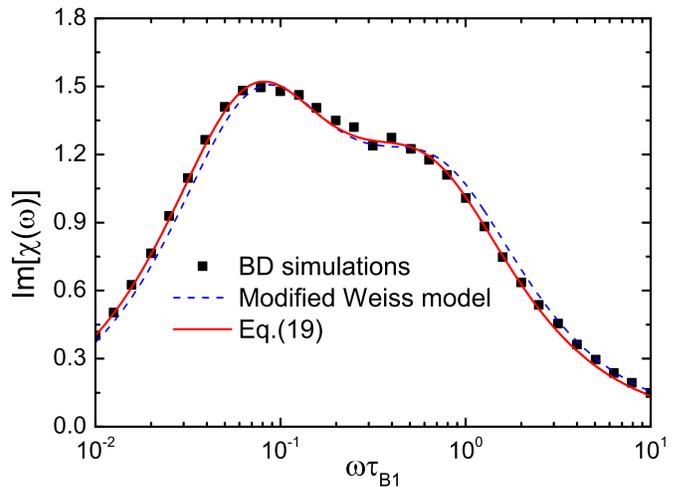}
 \hfill%
\caption {The imaginary part of DMS for a bidisperse ferrofluid. The frequency is measured in terms of the inverse of $\tau_{B1}$, which is the Brownian relaxation time for small particles (for large particles $\tau_{B2}=10\tau_{B1}$).  Symbols are from BD simulations for sample B in Ref.~\cite{CampMW:2018} (see Figure 5(b) in it).
The dashed blue line is from the modified Weiss model, while the solid red line is from my theoretical prediction based on Eq.~(19). Relevant material parameters  are provided in Table I of Ref.~\cite{CampMW:2018}.   By fitting Eq.~(19) with the results from BD simulations,  the values of $\tau_1$ and $\tau_2$ are determined as $\tau_1=1.137 \tau_{B1}$ and $\tau_2=1.078\tau_{B2}$, which seems quite reasonable~\cite{Fang:2020x2}.}
\end{figure}

To demonstrate the accuracy of my theoretic prediction, I plot in Fig.~3 the imaginary part of the DMS for a bidisperse ferrofluid sample.  This sample consists of two species (k=1, 2) of particles, with their sizes (no coating layer on every particle) and magnetic moments designed so that $\tau_{B2}/\tau_{B1}=10$ and
$\lambda_2=2\lambda_1=2$ (with $\lambda_k=\mu_0 \mu_k^2/4\pi d_k^3 k_B T$ characterizing the strength of DDI between k-particles). The number and volume fractions of small particles are $0.744$ and $0.181$, respectively.
As clearly shown in Fig.~3, the heuristic modified Weiss model agrees well with BD simulations, but shows observable deviations at both low and high frequencies.  Remarkably, my prediction based on Eq.~(19) is in nearly perfect agreement with BD simulations in the whole studied frequency range.  This is explained as follows.  According to my theory,
 the effects of inter-particle correlations are well separated into two parts. For the near-equilibrium behavior of ferrofluids,
the static correlations are captured by the factor $g_c = 1-\overline{\chi}_L/\overline{\chi}_0$, whereas the dynamic correlations are manifested by the ratios of the mean relaxation time $\tau_k$ to the Brownian relaxation time $\tau_{Bk}$.  In a manuscript being prepared by myself~\cite{Fang:2020x2},  it is proved that $\tau_k$ actually corresponds to the characteristic time describing the orientational self-diffusion of a tagged particle in the long time regime, during which it samples many configurations of surrounding particles.  As a consequence of the adiabatic approximation implied in deriving the mean field SE (and other dynamical density functional theories), it actually describes the stochastic dynamics of a representative dressed particle and $\tau_k/\tau_{Bk}$ characterizes the integrated (renormalization) effects of the neglected dynamic evolution of inter-particle correlations.

We usually have $\tau_k/\tau_{Bk} >1$ due to hydrodynamic and direct interactions.  Based on my theory and the observations from Fig.~3, the dynamic effect of inter-particle correlations can cause  enhanced dissipation at low frequencies but reduced dissipation at high frequencies.  This may have implications in applying ferrofluids to
magnetic therapy.  For the sample studied in Fig.~3,  the modified Weiss model well captures the effects of  static but not dynamic correlation,  while Eq.~(19) accounts for both.  This is why my theoretic prediction is even better.  Nevertheless, it is noted the DDI strength in all the bidispersed ferrofluid samples studied in Ref.~\cite{CampMW:2018} is only moderately strong. It remains of great interest to extensively investigate polydisperse samples with much stronger DDI (for which the modified Weiss model is even insufficient to capture the effect of static correlations) to check the accuracies of my theory.  Furthermore, it is extremely difficult (even for a monodisperse ferrofluid) to evaluate from first principles the value of $\tau_k$  due to the many-body and non-Markovian nature of the problem~\cite{Fang:2020x2}.  Nevertheless, the sensitivity of the DMS on $\tau_k$ may be exploited to determine it experimentally.

Finally, I determine $\overline{\tau}_B$, the characteristic relaxation time in the polydisperse MRE.  By requiring $\overline{\chi}_s(\omega)$ to match $\overline{\chi}(\omega)$ to the first order of $\omega$, we obtain
\begin{equation}
 \overline{\tau}_B = \sum_k p_k \chi^L_k \tau_k /\overline{\chi}_L.
\end{equation}
Notably, this expression is identical to the characteristic time defined by Ivanov et. al. based on their MMF1 theory~\cite{Ivanov:2016a}, although they do not distinguish $\tau_k$ from $\tau_{Bk}$.  Importantly, by comparing $\overline{\chi}_s(\omega)$ with $\overline{\chi}(\omega)$, we may determine the frequency cutoff or the critical time scale, $\tau_c$,  beyond which the polydisperse MRE is no longer valid.  On a time scale faster than $\tau_c$, we can not discard memory effects related  to inter-species coupling.  Then the instantaneous total magnetization is no longer a well-defined thermodynamic variable.

\section{Conclusions}
In conclusion, I have developed a nonperturbative and self-consistent dynamical mean field theory for mono- and polydisperse interacting ferrofluids.  I obtain  a generic magnetization relaxation equation and a general expression for the dynamic magnetic susceptibility, both of which are in simple and closed form.   My theory is expected to play a crucial role in studying ferrofluid dynamics, with important consequences on many areas of applications such as hyperthermia, drug delivery, and mechanical engineering. The general strategy presented here can be extended to other soft matter systems to construct dynamical mean field theories and obtain equations of motion for coarse-grained variables.

\section*{Acknowledgements}
I would like to thank Prof. Philip J. Camp for providing the data for Figure 1 and many helpful discussions. I appreciate the editor for his long waiting for my
revised manuscript.  I acknowledge the support from North China University of Water Resources and Electric Power via Grant No. 201803023.

\bibliographystyle{apsprb}

\begin{thebibliography}{18}

\email{afang1@gmail.com}
\bibitem{Rosen:1985}R. E. Rosenzweig, {\sl Ferrohydrodynamics}, Cambridge University Press, London, 1985.

\bibitem{Odenbach:2009}S. Odenbach (Ed.), {\sl Colloidal Magnetic Fluids: Basics, Development and Application
of Ferrofluids}, Springer, Berlin Heidelberg, 2009.


\bibitem{Ivanov:1992}Y. A. Buyevich and A. O. Ivanov, Physica A, 1992, {\bf 190}, 276.

\bibitem{Pshenichnikov:1995}A. F. Pshenichnikov, J. Magn. Magn. Mater., 1995, {\bf 145}, 319.

\bibitem{HukeLucke:2000}B. Huke, and M. L\"{u}cke, Phys. Rev. E, 2000, {\bf 62}, 6875.

\bibitem{Ivanov:2001} A. O. Ivanov, and O. B.  Kuznetsova,  Phys. Rev. E, 2001,  {\bf 64}, 041405.

\bibitem{HukeLucke:2004}B. Huke and M. L\"{u}cke, Rep. Prog. Phys., 2004, {\bf 67}, 1731.

\bibitem{Ivanov:2007}A. O. Ivanov, S.S. Kantorovich, E. N. Reznikov, C. Holm, A. F. Pshenichnikov, A.V. Lebedev, A. Chremos and P. J. Camp, Phys. Rev. E, 2007, {\bf 75}, 061405.

\bibitem{Ivanov:2017} A. Y. Solovyova, E. A. Elfimova, A. O. Ivanov and P. J. Camp, Phys. Rev. E, 2017, {\bf 96}, 052609.

\bibitem{Debye:1929}P. J. W. Debye, {\sl Polar Molecules}, Chemical Catalog Company, New York, 1929.

\bibitem{MRSh:1974}M. A. Martsenyuk, Y. L. Raikher and M. I. Shliomis, J. Exp. Theor. Phys., 1974, {\bf 38}, 413.

\bibitem{MullerLiu:2001}H. W. M\"{u}ller and M. Liu, Phys. Rev. E, 2001, {\bf 64}, 061405.

\bibitem{Felderhof:1999}B. U. Felderhof and H. J. Kroh, J. Chem. Phys., 1999, {\bf 110}, 7403.

\bibitem{Shliomis:2001c}M. I. Shliomis, Phys. Rev. E, 2001, {\bf 64}, 063501.

\bibitem{Fang:2020x1}A. Fang, ``The Dynamical Mean Field Model for Interacting Ferrofluids: I. Derivations for both homogeneous and inhomogeneous cases", in preparation.


\bibitem{Cichocki:1999}B. Cichocki, M. L. Ekiel-Je\`{z}ewska and E. Wajnryb,  J. Chem. Phys., 1999, {\bf 111}, 3265.

\bibitem{Nagele:2015}K. Makuch, M. Heinen, G. C. Abade and G. N\"{a}gele, Soft Matter, 2015, {\bf 11}, 5313.

\bibitem{Archer:2004}A. J. Archer and R. Evans, J. Chem. Phys., 2004, {\bf 121}, 4246.

\bibitem{Zubarev:1998}A. Y. Zubarev and  A. V. Yushkov, J. Exp. Theor. Phys., 1998, {\bf 87}, 484.

\bibitem{Ilg:2003}P. Ilg and S. Hess, Z. Naturforsch. A, 2003, {\bf 58}, 589.

\bibitem{Fang:2019b}A. Fang, Phys. Fluids, 2019, {\bf 31}, 122002.

\bibitem{Zwanzig:1961}R. Zwanzig, Phys. Rev., 1961, {\bf 124}, 983.
\bibitem{Grabert:1982}H. Grabert, {\sl Projection Operator Techniques in Nonequilibrium Statistical
Mechanics}, Springer Verlag, Berlin, 1982.

\bibitem{Blums:1997}E. Blums, A. Cebers and M. M. Maiorov,  {\sl  Magnetic Fluids},  Walter de Gruyter, Berlin, 1997.

\bibitem{Fang:2019a}A. Fang, ``Variational Approach to the Hydrodynamics of Interacting Ferrofluids", submitted.


\bibitem{CampMW:2018}A. O. Ivanov and P. C. Camp, Phys. Rev. E, 2018, {\bf 98}, 050602(R).

\bibitem{Ivanov:2016a}A. O. Ivanov, V. S. Zverev and S. S. Kantorovich, Soft Matter, 2016, {\bf 12}, 3507.

\bibitem{Note1}The near-equilibrium MMF1 theory presented in Ref.~\cite{Ivanov:2016a} is inequvalent to the model derived from our theory supplemented with the MMF1 EMOS in near-equilibrium regime.  To obtain the mean dipolar potential acting on a representative particle, their theory employs a Debye-Langevin equilibrium ODF for other particles.  Hence in their approximation ``other" particles are in a static configuration and not on the same footing as the representative particle.

\bibitem{Jones:2003}B. U. Felderhof and  R. B. Jones, J. Phys.: Condens. Matter, 2003,  {\bf 15}, 4011.


\bibitem{Ivanov:2016b}J. O. Sindt, P. J. Camp, S. S. Kantorovich, E. A. Elfimova and A. O. Ivanov, Phys. Rev. E, 2016, {\bf 93}, 063117.

\bibitem{Camp:2018}T. M. Batrudinov, Y. E. Nekhoroshkova, E. I. Paramonov, V. S. Zverev, E. A. Elfimova, A. O. Ivanov and P. J. Camp, Phys. Rev. E, 2018, {\bf 98}, 052602.

\bibitem{Raikher:1994}Y. L. Raikher and M. I. Shliomis,  Adv. Chem. Phys., 1994, {\bf 87}, 595.

\bibitem{Vroege:2003}B. H. Ern\'{e}, K. Butter, B. W. M. Kuipers and G. J. Vroege, Langmuir, 2003, {\bf 19}, 8218.

%


\bibitem{poly1:2016}A. O. Ivanov, S. S. Kantorovich, V. S. Zverev, E. A. Elfimova, A. V. Lebedev and A. F. Pshenichnikov, Phys. Chem. Chem. Phys., 2016, {\bf 18}, 18342.

\bibitem{poly2:2017}A.O. Ivanov, S. S. Kantorovich,  E. A. Elfimova, V. S. Zverev, J. O. Sindt and P. J.  Camp, J. Magn. Magn. Mater., 2017,  {\bf 431}, 141.

\bibitem{poly3:2018} A. O. Ivanov, S. S. Kantorovich, V. S. Zverev, A. V. Lebedev, A. F. Pshenichnikov and P. J. Camp, J. Magn. Magn. Mater., 2018, {\bf 459}, 252.


\bibitem{Fang:2020x2}A. Fang, ``The Dynamical Mean Field Model for Interacting Ferrofluids: II. The proper relaxation time and effects of dynamic correlation", in preparation.

%


\end{thebibliography}

\newpage

\end{document}